\newif\ifJFR
\JFRfalse
\newif\ifLNCS
\LNCStrue
\ifJFR
\documentclass{jfrarticle}
\usepackage{minted}\definecolor{vertfonce}{rgb}{0,.45,0}
\else
\ifLNCS  
\documentclass{llncs}
\usepackage{color}
\usepackage{a4}
\usepackage[margin=4cm]{geometry}
\else  
\documentclass{article}
\fi 
\fi 
\definecolor{pink}{rgb}{1,0.5,0.9}
\definecolor{vertfonce}{rgb}{0,.5,0}
\definecolor{brunfonce}{cmyk}{.3,.75,.75,.15}
\definecolor{beige}{cmyk}{0,.2,.7,0}

\newcommand{\bl}[1]{\textcolor{blue}{#1}}

\definecolor{vertfonce}{rgb}{0,.45,0}\newenvironment{minted}[1]{\begin{samepage}\begin{alltt}\color{vertfonce}}{\end{alltt}\end{samepage}}
\usepackage[utf8]{inputenc}
\usepackage[all]{xy}
\usepackage{qsymbols}
\usepackage{url}
\usepackage{amssymb}
\usepackage{wasysym}
\usepackage{hyperref}
\usepackage{alltt}

\newcommand{\Al}{\mbox{\footnotesize \textsf{A}}}
\newcommand{\Be}{\mbox{\footnotesize \textsf{B}}}

\newcommand{\Agda}{\textsf{Agda}}
\newcommand{\Coq}{\textsf{Coq}}
\newcommand{\Hol}{\textsf{HOL}}

\newcommand{\nat}{\ensuremath{\mathbb{N}}}

\newcommand{\nodA}{*++[o][F]{\Al}}

\newcommand{\nodB}{*++[o][F]{\Be}}

\newcommand{\fl}[1]{\ar@/^/[#1]^r \ar@/^/[d]^{d}}
\newcommand{\flr}[1]{\ar@[blue]@2@/^/[#1]^{\color{blue} r} \ar@/^/[d]^{d}}
\newcommand{\fld}[1]{\ar@/^/[#1]^r \ar@[blue]@2@/^/[d]^{\color{blue} d}}

\newcommand{\circu}{\raisebox{3.5pt}{\xymatrix{*+[o][F]{\tt u}}}}
\newcommand{\sgpe}{subgame perfect equilibrium}
\ifLNCS \else

\fi
\DeclareUnicodeCharacter{2192}{\ensuremath{\rightarrow}}
\DeclareUnicodeCharacter{27E8}{\ensuremath{\langle}}
\DeclareUnicodeCharacter{27E9}{\ensuremath{\rangle}}
\DeclareUnicodeCharacter{228E}{\ensuremath{\uplus}}
\DeclareUnicodeCharacter{2248}{\ensuremath{\approx}}
\DeclareUnicodeCharacter{2261}{\ensuremath{\equiv}}
\DeclareUnicodeCharacter{2E2}{\ensuremath{^s}}
\DeclareUnicodeCharacter{1D4D}{\ensuremath{^g}}
\DeclareUnicodeCharacter{2081}{\ensuremath{_1}}
\DeclareUnicodeCharacter{2082}{\ensuremath{_2}}
\DeclareUnicodeCharacter{25CB}{\ensuremath{\circ}}
\DeclareUnicodeCharacter{2193}{\ensuremath{\downarrow}}
\DeclareUnicodeCharacter{2191}{\ensuremath{\uparrow}}
\DeclareUnicodeCharacter{03C5}{\ensuremath{\upsilon}}
\DeclareUnicodeCharacter{24E4}{\ensuremath{\circu}}
\DeclareUnicodeCharacter{21CB}{\ensuremath{\rightleftharpoons}}
\DeclareUnicodeCharacter{263A}{\smiley}
\DeclareUnicodeCharacter{2115}{\ensuremath{\nat}}
\DeclareUnicodeCharacter{227D}{\ensuremath{\succcurlyeq}}
\DeclareUnicodeCharacter{226E}{\ensuremath{\nless}}

\usepackage{graphicx}
\newcommand{\textabstract}{Escalation in games is when agents keep playing
  forever.  Based on formal proofs we claim that \emph{if agents assume that resource
    are infinite, escalation is rational}.

\newcommand{\kwds}{extensive game, infinite game, sequential game, escalation
  coinduction, Agda, proof assistant, formal proof.}

\medskip

\noindent \textbf{Keywords:} \kwds}

\newcommand{\titre}{Extensive Infinite Games and Escalation,\\ an exercice in Agda}

\ifLNCS
\begin{document} 
\title{\titre}
\author{Pierre Lescanne}
\institute{University of Lyon, \'Ecole normale sup\'erieure de Lyon, CNRS (LIP), \\ 46 all\'ee
d'Italie, 69364 Lyon, France}

\maketitle

\noindent\rule{\textwidth}{.4pt}

\noindent\textabstract

\noindent\rule{\textwidth}{.4pt}
\else \ifJFR
\begin{document} 
\setcounter{page}{1}
\firstfoot{Journal of Formalized Reasoning  Vol.??, No.??, ??, Pages \pages.}
\runningfoot{Journal of Formalized Reasoning Vol.??, No.??, ??}
\title{\titre}
\author{Pierre Lescanne}{Pierre Lescanne\\\\
University of Lyon, \'Ecole normale sup\'erieure de Lyon, CNRS (LIP), \\ 46 all\'ee
d'Italie, 69364 Lyon, France}
\begin{abstract}
\textabstract
\end{abstract}
\maketitle
\else 
\begin{document}
\title{\titre}
\author{Pierre Lescanne\\\\
University of Lyon, \'Ecole normale sup\'erieure de Lyon, CNRS (LIP), \\ 46 all\'ee
d'Italie, 69364 Lyon, France}
\maketitle
\begin{abstract}
\textabstract
\end{abstract}
\fi %
\fi
\section{Introduction}
\label{sec:intro}

Escalation in games is the phenomenon where agents keep playing (or betting if
the game consists in bets) forever, leading to their ruin.  Since
Shubik\cite{Shubik:1971} people claim that such an attitude is not rational.  Based
on formal proofs we are able to refute such a claim and to say that \emph{if
  agents assume that resource are infinite, escalation is rational}.  Since our first
work\cite{DBLP:journals/corr/abs-0805-1798} which took place before the 2008
financial crisis, evidence\cite{bland18:_devel} show that stating the rationality of
escalation makes sense.  The only solution for avoiding escalation is then to assume
that resource are finite.

In previous
works\cite{DBLP:journals/corr/abs-1305-0101,DBLP:journals/corr/abs-1112-1185} we used
an approach based on \Coq{}\cite{boutillier:hal-01114602} and coinduction (a dual of
induction aimed at reasoning on infinite data structures\cite{DBLP:books/cu/12/0001R12}).  Especially
in\cite{DBLP:journals/jfrea/Lescanne18} we used dependent types together with
coinduction.  In this paper, we use coinduction in
\Agda{}\cite{DBLP:conf/tldi/Norell09}, because it allows a terse style closed to this
of mathematicians.  \Agda{} is a formal proof computer environment as well as a
dependently typed programming language.

Notice other works using proof assistants for proving properties of
agents. For instance, Stéphane Le~Roux proved the existence of Nash equilibria using \Coq{} and 
Isabelle\cite{DBLP:conf/tphol/Roux09,roux17:_formal_nash_coq_isabel}.  In a
somewhat connected area, Tobias Nipkow proved Arrows theorem in
\Hol{}\cite{DBLP:journals/jar/Nipkow09-2}.  \Agda{} code of this development are available on \href{https://github.com/PierreLescanne/DependentTypesForExtensiveGames-in-Agda}{GitHub}\,\footnote{\url{https://github.com/PierreLescanne/DependentTypesForExtensiveGames-in-Agda}}. 

\section{ Games and Strategy Profiles}
\label{sec:StPG}

Since we study game theory, lest us first define games. A \emph{game} is either a \emph{leaf} or a \emph{node}.
A leaf is a assignment to each agent of a \emph{Utility} (sometime called a
\emph{payoff}).  Note that the type of utility depends on the agent (dependent
type). A node contains two entities, put in a record: an \emph{agent} (the agent who
has the trait) and a function \emph{next} which tells the next positions to be played.
\pagebreak[4]
\begin{minted}{agda}
 mutual
  Game = ((a : Agent) → Utility a) ⊎ NodeG

  record NodeG : Set where
    coinductive
    field
      ag : Agent
      next : Choice → Game
\end{minted}
Notice the key word \textsf{coinductive} which shows that we deal with infinite
games.  The main concept in game theory is this of \emph{strategy profiles}.
Strategy profiles are like games with at each node a choice, which is the choice of
the agent who continues the game.  In \Agda{} the sum comes with to unctions
\textsf{inj₁} and \textsf{inj₂}.  In our case, if \textsf{u} is a utility assignment
of type \textsf{((a : Agent) → Utility a)} then \textsf{inj₁ u} is a \textsf{Game}
and \textsf{n} is a \textsf{NodeG} then \textsf{inj₂ n} is a \textsf{Game}.  Strategy
profiles are abbreviated \textsf{StratProf}.
\begin{minted}{agda}
mutual
  StratProf = ((a : Agent) → Utility a) ⊎ NodeS

  record NodeS : Set where
    coinductive
    field
      ag : Agent
      next : Choice → StratProf
      ch : Choice
\end{minted}
We can define the underlying game of a strategy profile
\begin{minted}{agda}
game : (s : StratProf) → Game
game (inj₁ u) = inj₁ u
game (inj₂ n) = inj₂ (gameN n) where
  gameN : NodeS → NodeG
  NodeG.ag (gameN n) = ag n
  NodeG.next (gameN n) c = game (next n c)  
\end{minted}
The underlying game of a leaf (strategy profile) is the same utility assignment, i.e., a leaf (game). For nodes, games are attributed corecursively. 
Now let us look at another concept. 
Given two strategy profiles, one may wonder whether they have the same
underlying game. This is given by the binary relation \_≈ˢᵍ\_.
\begin{minted}{agda}
mutual
  data _≈ˢᵍ_ : StratProf → StratProf → Set where
    ≈ˢᵍLeaf : {u : (a : Agent) → Utility a} → inj₁ u ≈ˢᵍ inj₁ u
    ≈ˢᵍNode : {n n' : NodeS} → n ○≈ˢᵍ n' → inj₂ n ≈ˢᵍ inj₂ n'

  record _○≈ˢᵍ_ (n n' : NodeS) : Set where
    coinductive
    field
      is○≈ˢᵍ : ag n ≡ ag n' → ((c : Choice) → next n c ≈ˢᵍ next n' c)
\end{minted}
A leaf has the same game as itself, two nodes have the same game if all their
``next'' strategy profiles have the same games.  Notice that we use the symbol
$\circ$ for concepts associated with \textsf{NodeS}, when the concept without $\circ$
is associated with \textsf{StratProf}.  Given a strategy profile, we may want to
compute the utility of an agent.  This assumes that the path that follows the choices
of the agents leads to a leaf.  A strategy profile~\textsf{s} with such a property is
said \textsf{convergent}, written $\downarrow$~\textsf{s}.  This is defined as
follows:
\begin{minted}{agda}
mutual
  data ↓ : StratProf → Set where
    ↓Leaf : {u : (a : Agent) → Utility a} → ↓ (inj₁ u)
    ↓Node : {n : NodeS} → ○↓ n → ↓ (inj₂ n)

  record ○↓ (n : NodeS) : Set where
    inductive
    field
      is○↓ : ↓ (next n (ch n))  
    \end{minted}
Notice that  not all the strategy profile are convergent, for instance the strategy profile \textsf{AcBc} of Section~\ref{sec:two} is not convergent.

We define the utility assignment \circu{} of a convergent strategy profile. 
\circu{} takes two parameters: a strategy profile \textsf{s} and a proof
that \textsf{s} is convergent.
\begin{minted}{agda}
  ⓤ : (s : StratProf) → (↓ s) → (a : Agent) → Utility a
  ⓤ (inj₁ u) ↓Leaf = u
  ⓤ (inj₂ n) (↓Node p) = ○ⓤ n p

  ○ⓤ : (n : NodeS) → (○↓ n)  → (a : Agent) → Utility a
  ○ⓤ n p = ⓤ (next n (ch n)) (is○↓ p)
\end{minted}
\emph{Subgame perfect equilibria} are very interesting strategy profiles.  They are
strategy profiles in which the choices of the agents are the best.  A leaf is always a
\sgpe{}. A~node is a \sgpe{} if the next strategy profile for the choice of the agent
is convergent and is a \sgpe{}, if for any other node which has the same game and
whose next strategy profile is also convergent and is a \sgpe{}, the utility of the
agent of  the given node is not less than the utility of the agent of this other node.
This is defined formally in \textsf{Agda} as follows, where we use
$\rightleftharpoons$~\textsf{s} to tell that \textsf{s} is a \sgpe.
\begin{minted}{agda}
data ⇋_ : StratProf → Set where
  ⇋Leaf : {u : (a : Agent) → Utility a} → ⇋ inj₁ u
  ⇋Node : {n n' : NodeS} →
    n ○≈ˢᵍ n' →
     ⇋ (next n (ch n)) →
     ⇋ (next n' (ch n')) →
    (p : ↓ (next n (ch n))) → (p' : ↓ (next n' (ch n'))) →
    (ⓤ (next n (ch n)) p (ag n)) ≮ (ⓤ (next n' (ch n')) p' (ag n)) →
     ⇋ inj₂ n
\end{minted}

\section{Escalation}
\label{sec:esca}

We are now interested in strategy profile leading to escalation. 

\subsection{Good strategy profile}
\label{sec:good}

A first property toward escalation is what we call \emph{goodness}.  A strategy profile is
\emph{good} if at each node, there is a \sgpe{} with the same game and the same
choice.
\begin{minted}{agda}
mutual
  data ☺_ : (s : StratProf) → Set where
    ☺Node :  {n : NodeS} → ○☺ n → ☺ (inj₂ n)
  record ○☺_ (n : NodeS) : Set where
    coinductive
    field
      is○☺ : (n' : NodeS) → ⇋ (inj₂ n') → n ○≈ˢᵍ n' → ch n ≡ ch n' →
        ☺ (next n (ch n))
\end{minted}
In other words, this strategy profile is not itself a \sgpe, in particular, it can be
non convergent, but each of its choices is dictated by a \sgpe.  Goodness can be considered as \emph{rationality} in the choices of the agents.  Reader may notice that goodness is of interest only in infinite games, because in a finite game, there is no difference between a good strategy and a \sgpe.

\subsection{Divergent strategy profile}
\label{sec:div}

Another property of strategy profiles is \emph{divergence}. In a divergent strategy
profile, if one follows the choices of the agents, one never gets to a leaf, but, on
the opposite, one runs forever. A divergent strategy profile is written
$\uparrow$~\textsf{s}.  The formal definition in \textsf{Agda} of divergence looks like this of
convergence, but the test for divergence is based on a coinductive record and never
hits a leaf, therefore there is no \bl{$\uparrow$\textsf{Leaf}} case.
\begin{minted}{agda}
mutual
  data ↑_ : StratProf → Set where
    ↑Node :  {n : NodeS} → ○↑ n → ↑ (inj₂ n)

  record ○↑ (n : NodeS) : Set where
    coinductive
    field
      is○↑ : ↑ (next n (ch n))
\end{minted}

An \emph{escalation} is a strategy profile which is both \emph{good} and \emph{divergent}.

\section{Strategies with two agents and two choices}
\label{sec:two}

To build escalating strategy profiles, we consider the case of two agents
\textsf{Alice} and \textsf{Bob} and two choices \textsf{down} and \textsf{right}. 
\begin{minted}{agda}
data AliceBob : Set where
  Alice Bob : AliceBob  
\end{minted}
\begin{minted}{agda}
data DorR : Set where
  down right : DorR  
\end{minted}
We take the natural numbers $\nat$ as utility\,\footnote{We could have taken a utility
  with only two values, but we feel that the reader is more acquainted with natural
  numbers for utilities.} for both agents\,\footnote{In this case, the type of utility
  does not depend on the agent.} and for the $\nless$ relation we take the
$\succcurlyeq$ relation defined as:
\begin{minted}{agda}
data _≽_ : ℕ → ℕ → Set where
  z≽z : zero ≽ zero
  s≽z : {n : ℕ} → suc n ≽ zero
  s≽s : {n m : ℕ} → n ≽ m → suc n ≽ suc m
\end{minted}

A utility assignment is for instance this which assigns $1$ to \textsf{Alice} and
$0$ to \textsf{Bob}:
\begin{minted}{agda}
uA1B0 : AliceBob → ℕ
uA1B0 Alice = 1
uA1B0 Bob = 0
\end{minted}
from which we can build a leaf strategy profile:
\begin{minted}{agda}
A1B0 : StratProf
A1B0 = inj₁ uA1B0
\end{minted}
which is convergent.
\begin{minted}{agda}
↓A1B0 : ↓ A1B0
↓A1B0 = ↓Leaf
\end{minted}
From the utility assignment which assigns $0$ to \textsf{Alice} and $1$ to
\textsf{Bob} on can build the convergent strategy profile \texttt{A0B1}.

Moreover, we build an infinite strategy \textsf{AcBs}, in which \textsf{Alice}
continues always and \textsf{Bob} stops always:
\begin{minted}{agda}
mutual 
  AcBs : StratProf
  AcBs = inj₂ ○AcBs

  ○AcBs : NodeS
  ag ○AcBs  = Alice
  ch ○AcBs  = right
  next ○AcBs down = A0B1
  next ○AcBs right =  BsAc
  
  BsAc : StratProf
  BsAc = inj₂ ○BsAc

  ○BsAc : NodeS
  ag ○BsAc = Bob
  ch ○BsAc = down
  next ○BsAc down = A1B0
  next ○BsAc right = AcBs
\end{minted}
We notices that by mutual co-recursion, \textsf{AcBs} is defined together with an infinite
strategy profile \textsf{BsAc} which starts with a node of which \textsf{Bob} is the
agent.  Those strategies are like infinite combs. 
 
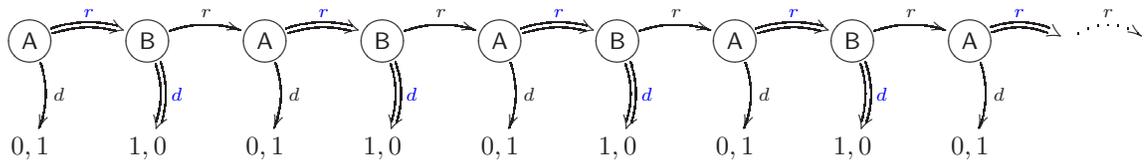
\begin{displaymath}
    \xymatrix{
      \nodA\flr{r}&\nodB\fld{r}&\nodA\flr{r}&\nodB\fld{r}&\nodA\flr{r}&\nodB\fld{r}%
      &\nodA\flr{r} &\nodB\fld{r}&\nodA\flr{r} &\ar@{.>}@/^/[r]^r&\\
      0,1&1,0&0,1&1,0&0,1&1,0&0,1&1,0& 0,1&&}
  \end{displaymath}
With \textsf{down} one reaches always a leaf
and with \textsf{right} one goes always to a new strategy profile, which is a node.  There is a
variant of the node $\circ$\textsf{AcBs}, in which the first choice of \textsf{Alice}
is \textsf{down} instead of \textsf{right}.
\begin{minted}{agda}
Var○AcBs : NodeS
ag Var○AcBs  = Alice
ch Var○AcBs  = down
next Var○AcBs down = A0B1
next Var○AcBs right = BsAc
\end{minted}
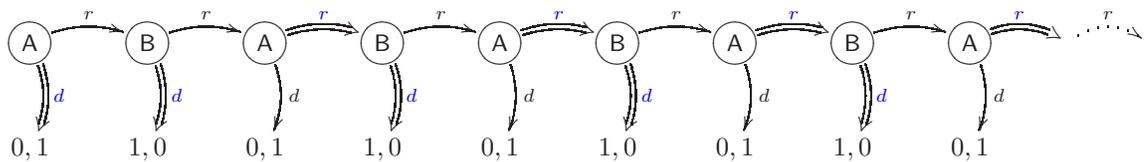
\begin{displaymath}
    \xymatrix{
      \nodA\fld{r}&\nodB\fld{r}&\nodA\flr{r}&\nodB\fld{r}&\nodA\flr{r}&\nodB\fld{r}%
      &\nodA\flr{r} &\nodB\fld{r}&\nodA\flr{r} &\ar@{.>}@/^/[r]^r&\\
      0,1&1,0&0,1&1,0&0,1&1,0&0,1&1,0& 0,1&&}
\end{displaymath}
We prove that $\circ$\textsf{AcBs} and \textsf{Var}$\circ$\textsf{AcBs} have the same
game.  Likewise we prove that \textsf{AcBs} is convergent i.e.,
$\downarrow~$\textsf{AcBs}.  Those two facts are key steps in the proof that
\textsf{AcBs} is subgame prefect equilibrium i.e., that
\ensuremath{\rightleftharpoons}~\textsf{AcBs}.

On the same paradigm we built a strategy profile \textsf{AsBc} in which \textsf{A}
stops and \textsf{B} continues and which is proved to be convergent and to be a subgame
perfect equilibrium.

We also build a strategy profile in which \textsf{A} and \textsf{B} both continue.
\begin{minted}{agda}
mutual 
  AcBc : StratProf
  AcBc = inj₂ ○AcBc

  ○AcBc : NodeS
  ag ○AcBc  = Alice
  ch ○AcBc  = right
  next ○AcBc down = A0B1
  next ○AcBc right =  BcAc
  
  BcAc : StratProf
  BcAc = inj₂  ○BcAc

  ○BcAc : NodeS
  ag ○BcAc = Bob
  ch ○BcAc = right
  next ○BcAc down = A1B0
  next ○BcAc right = AcBc  
\end{minted}
\textsf{AcBs}, \textsf{AcBc} and \textsf{AsBc} have the same game.  Unlike \textsf{AcBs} and \textsf{AsBc}, the strategy profile \textsf{AcBc} is divergent, i.e., \textsf{↑\,AcBc}.  Moreover \textsf{AcBc} is good which means \textsf{☺AcBc}.

\section{Conclusion}
\label{sec:conc}

Since \textsf{AcBc} is good and divergent, \textsf{AcBc} \textbf{is an
  escalation}. Hence we proved formally the claim of the introduction, namely
\emph{if agents assume that resource are infinite, escalation is rational}.

In the current implementation, the type of choices is the same for all the
agents. However, one may imagine that this type may depend on the agents. Making the
type of choices depending on the agents is object of the current investigation.


\end{document}